\documentclass{svjour2}
\usepackage{color}
\usepackage{graphicx}
\usepackage{amsthm,amsmath, amssymb,mathtools}
\usepackage{slashbox}
\usepackage{enumerate}
\usepackage{tabularx}

\usepackage{hyperref}
\hypersetup{
    colorlinks=true,
    linkcolor=blue,
    citecolor=blue,
    filecolor=magenta,      
    urlcolor=cyan,
}
\newcommand{\be}{\begin{equation}}
\newcommand{\ee}{\end{equation}}
\newcommand{\bea}{\begin{align}}
\newcommand{\eea}{\end{align}}

\begin{document}
\title{High-Activity Expansion for the Columnar Phase of the Hard
Rectangle Gas}
\author{Trisha Nath \and Joyjit Kundu \and R. Rajesh}
      
\institute{
Trisha Nath\and Joyjit Kundu \and R. Rajesh \at The Institute of Mathematical Sciences, C.I.T. Campus,
Taramani, Chennai 600113, India\\
\email{trishan@imsc.res.in, joyjit@imsc.res.in, rrajesh@imsc.res.in}
}

\date{\today}

\maketitle

\begin{abstract}

We study a system of monodispersed hard rectangles of size $m \times d$, 
where $d\geq m$ on a two dimensional square lattice. For large enough
aspect ratio, the system is known to
undergo three entropy driven phase transitions with increasing
activity $z$: first from disordered to nematic, second from 
nematic to columnar and third from  columnar to sublattice phases. We study
the nematic-columnar transition by developing a high-activity
expansion in integer powers of $z^{-1/d}$ 
for the columnar phase in a model where the rectangles are
allowed to orient only in one direction. By deriving the exact
expression for the first $d+2$ terms in the expansion, we obtain lower
bounds for the critical density and activity. For $m$, $k\gg 1$, these bounds 
decrease with increasing $k$ and decreasing $m$.
\end{abstract}


\section{\label{I}Introduction}

A system of long rods in three dimensions with only  excluded volume interactions is known to exhibit
a density-driven phase  transition from a disordered isotropic phase to an
orientationally ordered nematic phase~\cite{onsager,flory,z63,degennes,vroege}.
Further increase in density may result in a smectic phase with orientational order
and partial translational order, and a solid 
phase~\cite{frenkel,bolhuis}. In two dimensional continuum 
space, the system undergoes a Berezinskii-Kosterlitz-Thouless 
transition~\cite{b71,b72,kt73} from a 
low-density phase with exponential decay of correlations 
to a high-density phase where the correlations decay as a power 
law~\cite{straley1971,frenkel1985,khandkar2005,vink2009}.

The corresponding problem has also been studied on lattices where
the rods orient only along the lattice directions, and thus the number of
allowed orientations is finite.
In this case, it
may be heuristically argued that the fully packed phase has no 
orientational order~\cite{degennes,deepak}, making it uncertain whether
a pure lattice model may ever exhibit a nematic phase~\cite{degennes}. 
There has been a renewed interest in this problem after it was  convincingly
demonstrated numerically that a system of  hard rods on a square lattice 
exhibits a nematic phase for large enough aspect ratio~\cite{deepak}.
Below, we summarize the known results for lattice models of hard 
rods and rectangles.

Consider a mono-dispersed system of hard rectangles of
size $m \times d$ on a square lattice, where $k=d/m$ is the aspect
ratio, and each rectangle may orient along one of two directions
-- horizontal or vertical. No two rectangles may overlap
(see Sec.~\ref{II} for a more precise definition). When $m=1$ (hard rods),
the system has been shown to undergo two density-driven transitions:
a low-density isotropic--nematic 
transition shown numerically for $k \geq 7$~\cite{deepak} and rigorously for 
$k \gg 1$~\cite{dg13}, and
a second high-density  nematic--disordered transition that has been
shown numerically~\cite{krds12,kundu2}.
While the first transition is in the Ising universality 
class~\cite{flr08,flr08a,fernandez2008c,linares2008,fischer2009}, there is no 
clear-cut
evidence for the second transition belonging to any known
universality class~\cite{kundu2}. The $m=1$ model may be solved exactly on 
a tree-like lattice where the system undergoes an isotropic--nematic
transition for $k \geq 4$, but the second transition is absent~\cite{drs11}, though an
exact solution for rods with soft repulsive interactions on the same lattice shows
two transitions~\cite{kr13}. The only other exact result is 
for $m=1$ and $k=2$ (dimers), 
where
the absence of a transition for any density may be 
proved~\cite{Heilmann1970,Kunz1970,Gruber1971,lieb1972}. The system with
$m >1$ has also been studied.
When $k=1$, and 
$m\geq 2$, the model reduces to the well-studied hard square 
model~\cite{bellemans_nigam1nn,bellemans_nigam3nn,ree,landau,nisbet1,lafuente2003phase,ramola,schmidt,slotte,kinzel,amar} which
undergoes a density-driven transition from a disordered to a columnar phase. 
The transition is continuous for $m=2$ ~\cite{fernandes2007,feng,zhi2} 
and first order for $m = 3$~\cite{fernandes2007}. The columnar phase breaks
translational symmetry only in one lattice direction.
When $m\geq 2$ and $k\geq 7$, the system undergoes three density-driven
transitions: first an isotropic-nematic transition, second a nematic-columnar 
transition, and third a columnar-sublattice transition~\cite{kundu3,kr14b}. Here, 
the columnar phase
breaks translational symmetry in the direction perpendicular to the nematic
orientation. For $2 \leq k <7$, and $m\geq 3$, the system undergoes 
isotropic-columnar and columnar-sublattice transitions with increasing
density. When $m=2$, the system undergoes a direct isotropic-sublattice
transition for $k=2,3$~\cite{kundu3}.

In this paper, we focus on the nematic-columnar transition. 
When $m=2$, the critical density $\rho_c$ (the fraction of occupied lattice sites)  
for this transition was numerically determined for $d$ up to $48$.
By extrapolating to $d\to \infty$, it was shown that 
$\rho_c\approx  0.727 + 0.226 k^{-1} + O(k^{-2})$~\cite{kr14b}, implying the
existence of the columnar phase for infinitely long rectangles. It is 
difficult to study systems with larger $m$ using Monte Carlo simulations 
as it becomes difficult
to equilibrate the systems at large densities, restricting the
numerical study of large $k$ to $m=2$. The nematic-columnar
transition has also been studied analytically within a 
Bethe approximation~\cite{kundu3}. While this calculation
qualitatively reproduces the above numerical result, the approximations
involved are ad-hoc with no clear systematic procedure of improving 
the results. In 
addition, the limit $m\to \infty$ keeping $k=d/m$ fixed, corresponding
to a system of oriented rectangles in the continuum,
is unattainable. 
As of now, unlike the nematic phase, there exists no rigorous proof for the 
existence of the columnar phase.

High-activity expansions are a systematic and more rigorous way of studying
the effect of fluctuations in the ordered state. In the standard Mayer
expansion, the high-activity expansion is in integer powers of $z^{-1}$, where
$z$ is the activity or fugacity~\cite{gaunt_fisher}. However, columnar phases 
possess a sliding instability resulting in the expansion being in 
fractional powers of $z^{-1}$~\cite{bellemans_nigam1nn}. The expansion
was carried out to $O(z^{-3/2})$ for the hard square gas ($m=2$, $d=2$) 
recently~\cite{ramola}, and the formalism was applied to hard core lattice
gas models with first four next nearest neighbor exclusion, where the
high-density phase is columnar~\cite{trisha}.

In this paper, we generalize the calculations of the hard square gas in
Ref.~\cite{ramola} to derive the high-activity expansion for the columnar
phase of the hard rectangle gas. To do so, we study a simpler model
where rectangles may orient only in the horizontal direction. We
justify this simplification by arguing that near the nematic-columnar
transition, there are only a few rectangles with orientation perpendicular to
the nematic orientation. In addition, we also show that the
nematic-columnar transition densities are the same for large $k$
(within numerical error), whether both orientations or only one orientation 
is allowed.  
We show that the high-activity expansion is
in powers of $z^{-1/d}$, where $z$ is the activity or fugacity. The
exact expressions for the
first $d+2$ terms in the expansion of free energy and densities
are derived. Truncating the expansions for the densities at this
order, we obtain estimates (lower bounds) for the critical densities
for the nematic-columnar transition. For large $m$ and $k$, these estimates are shown to
decrease with increasing $k$ and decreasing $m$.

The rest of the paper is organized as follows.
Section~\ref{II} contains a definition of the model and the
justification for studying a model of hard rectangles with only
horizontal orientation. The 
high-activity expansion for the free energy of $m\times d$ rectangles 
is derived in Sec.~\ref{III}. In Sec.~\ref{IV}, we derive the
high-activity expansion for the occupation densities of the different
rows.  The critical densities, and activities are estimated from these expansions.
Section~\ref{V} contains a discussion of the results and some possible 
extensions of the problem.

\section{\label{II}Model and Preliminaries}

Consider a square lattice of size $L \times L$ with periodic boundary 
conditions. We consider a system of monodispersed hard rectangles of size 
$m \times d$, where $k=d/m$ is the aspect ratio, and $d \geq m$. 
A horizontal (vertical)
rectangle occupies $d$ ($m$) 
consecutive lattice sites along the $x$-direction and $m$ ($d$) 
consecutive lattice sites along the $y$-direction. No lattice site may 
be occupied 
by more than one rectangle. The grand canonical partition function for 
the system is 
\be
\overline{\mathcal{L}}(z_h,z_v)=\sum \limits_{n_h,n_v} \overline{C}(n_h,n_v) z_h^{n_h}z_v^{n_v},
\label{eq:grandcanonical}
\ee
where $\overline{C}(n_v,n_h)$ is the number of valid configurations with 
$n_h$ horizontal rectangles  and $n_v$ 
vertical rectangles, and $z_h$ and $z_v$ 
are the corresponding activities. 

\begin{figure}
\centering
\includegraphics[width=\columnwidth]{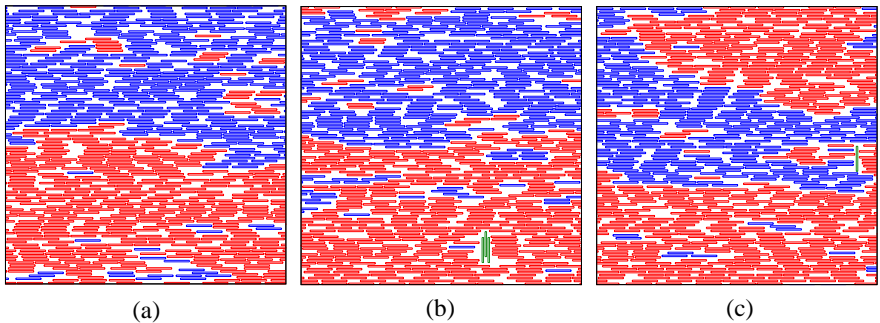}
\caption{Snapshots of a system of rectangles of size $2 \times 18 (k=9)$ at
Monte Carlo time 
steps 
(a) $7 \times 10^7$,
(b) $8 \times 10^7$, and
(c) $9 \times 10^7$.
Red 
and blue- are horizontal rectangles with heads at the even and odd rows 
respectively and green are vertical rectangles. 
 }
\label{fig:fig01}
\end{figure}

In the nematic and columnar phases, the orientational symmetry is 
broken, and thus the majority of rectangles are either horizontal or 
vertical. Typical snapshots of the system ($m=2$, $d=18$) at equilibrium near the
nematic-columnar transition are shown in Fig.~\ref{fig:fig01}. 
The system has nearly complete orientational order, and one may ignore the
effects of the rectangles with perpendicular orientation.
To study the nematic--columnar phase transition, it is thus more 
convenient to study a system in which all rectangles are horizontal. 
Thus, we set $z_v=0$ in (\ref{eq:grandcanonical}), disallowing vertical 
rectangles. 

\begin{figure}
\centering
\includegraphics[width=0.9 \columnwidth]{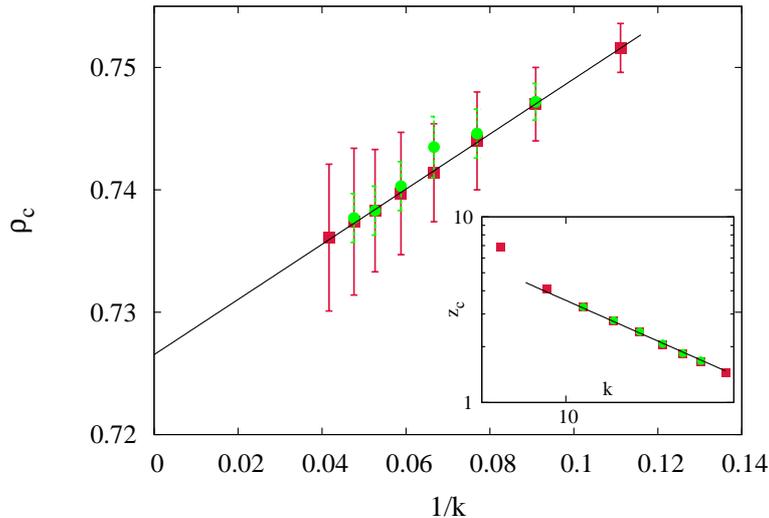}
\caption{The critical density $\rho_c$ for the nematic-columnar
transition for the model with only horizontal rectangles (circles) and
the model with both horizontal and vertical rectangles (squares) from Monte Carlo simulation. The
data for the latter is from Ref.~\cite{kr14b}. The straight line is
$0.727+0.226 k^{-1}$. Inset: The corresponding 
critical activity $z_c$ from Monte Carlo simulation. The straight line is
$35.5 k^{-1}$.
}
\label{fig:fig02}
\end{figure}

Further justification of this simplification may be obtained by
comparing the critical density $\rho_c$ for the nematic-columnar
transition for the model of rectangles with both orientations and the model of
rectangles with only horizontal orientation. $\rho_c$ for the 
model with both orientations allowed
was numerically obtained for $m=2$ as $\rho_c \approx
0.727 + 0.226 k^{-1}$ for $k \gg 1$~\cite{kr14b}. Here, we obtain
$\rho_c$ for the model restricted to horizontal rectangles from
Monte Carlo simulations. The details of the algorithm and parameters
are as in Refs.~\cite{kundu3,kr14b}. $\rho_c$ is obtained for
$m=2$ by the intersection point of the Binder cumulant for three
different lattice sizes, and is shown in Fig~\ref{fig:fig02}. We
obtain $\rho_c \approx 0.727 + 0.226 k^{-1}$ for $k \gg 1$,
numerically indistinguishable from that for the model with both
horizontal and vertical rectangles. 
We thus conclude that the
simplified model is well-suited for studying the nematic-columnar
transition.
However,  it is possible that the two models
have qualitatively different phenomenology  for small $k$
when the nematic phase is absent for the model with both orientations allowed
(also see Sec.~\ref{V} for more
discussion of this point). 

\begin{figure}
\centering
\includegraphics[width=0.9 \columnwidth]{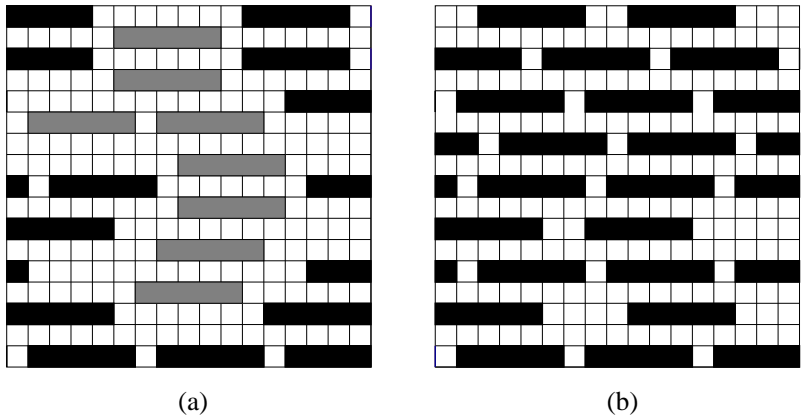}
\caption{A schematic diagram illustrating the two phases for $2 \times
6$ rectangles.  (a) The low activity nematic phase where some of the
heads (bottom left corner) of the rectangles are on odd rows (black) 
and some of the heads are on even rows (grey).
(b) The high activity columnar phase where most of the heads of
rectangles are either on even or odd rows (odd in the example shown).
}
\label{fig:fig03}
\end{figure}

Let the bottom left corner of a rectangle be called its head. In the 
nematic phase, each row on an average contains equal number of heads of 
rectangles. In the columnar phase, this symmetry is broken. An example
illustrating the two phases for $2 \times 6$ rectangles is shown in 
Fig.~\ref{fig:fig03}. To quantify 
the nematic--columnar transition, we assign to the $i^{th}$ row a 
label $(i \mod m)+1$, 
such that the labels are $1, \ldots, m$. In the columnar phase, 
majority of the rectangles have their heads on one of the $m$ types of rows.
The grand canonical partition function for the model with only
horizontal rectangles is then
\be
\label{gps}
\mathcal{L}(\{z_i\})=\sum_{n_1, \ldots, n_m} C(n_1, \ldots, n_m) 
\prod_{i=1}^{m} z_i^{n_i},
\ee
where $C(n_1, \ldots, n_m)$ is the number of configurations with $n_i$ 
rectangles whose heads are on rows with label $i$, and $z_i$'s are the 
corresponding activities. For large activities, the system will be in the columnar
phase, and undergoes a transition to the nematic phase as the activities
are decreased.

The free energy of the system in the thermodynamic limit is 
\be
\label{free}
f(z_1,\ldots,z_m)=\lim_{N\to \infty}- \frac{1}{N}\ln
\mathcal{L}(z_1,\ldots,z_m),
\ee
where $N=L^2$ is the total number of lattice sites. The density of
occupied sites $\rho$ is then given by
\be
\label{density}
\rho(z)=-m d z\frac{d}{ d z}  f(z,\ldots,z).
\ee
The aim of the paper is to determine the free energy $f$ and density
$\rho$ as a perturbation series in inverse powers of the activity.

\section{\label{III} High activity expansion for the free energy}

The perturbation series  for the free energy will be in
powers of $z^{-1/d}$ [see Sec.~\ref{sI} and~\eqref{eq:f0}] 
rather than the usual Mayer expansion in
integer powers of $z^{-1}$ because of the ordered columnar phase
having a sliding instability~\cite{bellemans_nigam1nn,ramola}. Suppose a 
vacancy is created in a fully
ordered columnar phase at full packing by removing a rectangle. The $d$
consecutive empty sites that are created may now be broken into $d$
fractional vacancies without any loss of entropy by sliding sets of
rectangles in the horizontal direction. This breaking up of vacancies into 
fractional vacancies leads to fractional powers of $z^{-1}$ appearing in 
the perturbation expansion.

To set up a perturbation expansion about the ordered columnar 
state, it is convenient to
choose one of the activities to be large and treat the other activities as 
small parameters:
\begin{align}
z_1= z_o,\label{z_def1}\\
z_2=\ldots=z_m=z_e,\label{z_def2}
\end{align}
where $z_e \ll z_o$. The notation is such that when $m=2$, $o$ indicates
odd rows and $e$ indicates even rows. Once the perturbation expansion
is obtained, $z_o$ and $z_e$ are equated to $z$ to obtain the
high-activity expansion.

\begin{figure}
\centering
\includegraphics[width=\columnwidth]{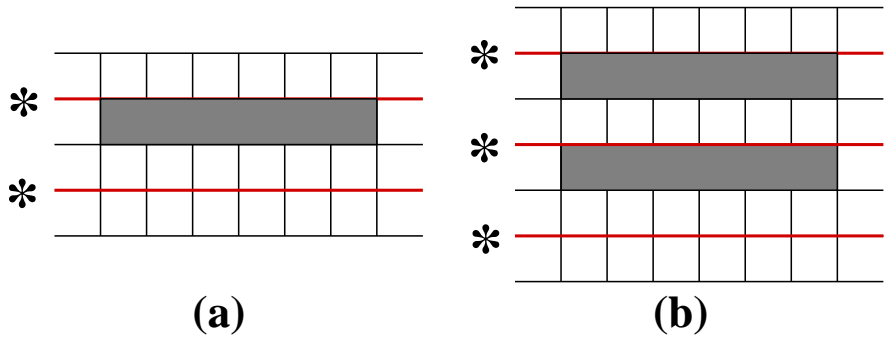}
\caption{A schematic diagram illustrating clusters.  (a) One rectangle on 
an even row results in reducing the maximal occupancy of two
odd rows (red lines labeled \textasteriskcentered) by one. (b) Two 
rectangles on even rows one directly above the other results in
reducing the maximal occupancy of three odd rows (red lines
labeled \textasteriskcentered) by one.
}
\label{fig:fig04}
\end{figure}

In the completely ordered state, heads of all the rectangles are in rows with
label $1$ [see Fig.~\ref{fig:fig03}(b)]. In the perturbation expansion, we 
refer to rectangles whose
heads are in rows with a label different from $1$ as defects. For
systems with sliding instability, the perturbation expansion is not in
terms of number of defects, but rather in terms of clusters of
defects~\cite{bellemans_nigam1nn,ramola}. We illustrate this for the
case $m=2$. Consider a
single defect on a row with label $2$ [see Fig.~\ref{fig:fig04}(a)]. This 
defect results in one or more rectangles removed from each of the
two rows denoted by $\ast$ in Fig.~\ref{fig:fig04}(a), both with label
$1$, and therefore has leading
weight $z^{-1}$. Now consider two defects both on rows with label $2$
but one directly above the other [see Fig.~\ref{fig:fig04}(b)]. 
This cluster of defects results in one or more  rectangles
removed from each of the three rows denoted by $\ast$ in
Fig.~\ref{fig:fig04}(b), all three with label $1$,
and therefore still has leading weight $z^{-1}$. It is easy to see that a 
similar defect cluster of
arbitrary size will have leading weight $z^{-1}$. When $m>2$,
defect-clusters may have defects with different labels (as defined
below). For such clusters, 
it is again 
possible that defect-clusters of different sizes
also have leading weight $z^{-1}$. 

\begin{figure}
\centering
\includegraphics[width=0.7 \columnwidth]{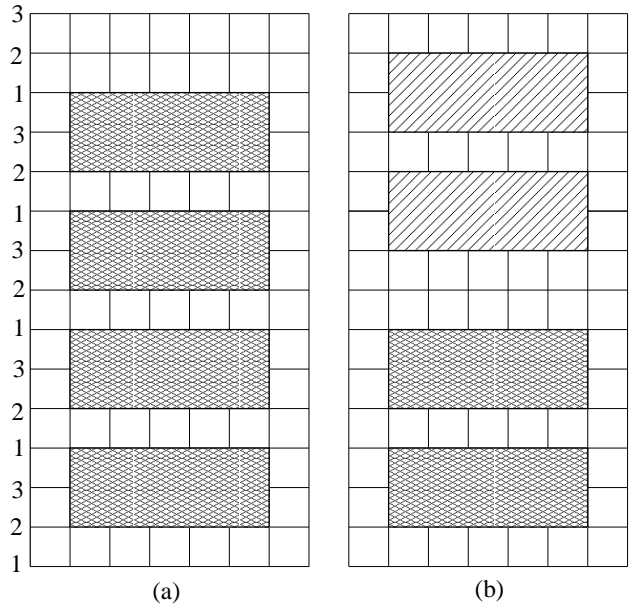}
\caption{An example of a single cluster of size $4$ made up 
of (a) 4 defects of label 2 and (b) 2 defects of label 2 and
2 defects of label 3. The example is for rectangles of size $3 \times 6$.
}
\label{fig:fig05}
\end{figure}

We define a single cluster of defects more precisely. Let a
sub-cluster of size $n$ and label $j$ denote a set of $n$ rectangles
of label $j$, directly on top of each other such that the long sides
are parallel and the short sides are aligned. 
A single cluster of
defects is made up of sub-clusters of label $2, 3, \ldots m$,
such that the labels are in ascending order and the gap between two
sub-clusters is the minimum possible. 
Examples of single clusters of
size $4$ for $3 \times 6$ rectangles are shown in Fig.~\ref{fig:fig05}. 
In  Fig.~\ref{fig:fig05}(a), the cluster is made up of a single
sub-cluster of label $2$ while in Fig.~\ref{fig:fig05}(b), the
cluster is made up of one sub-cluster of size 2 of label $2$ and
one sub-cluster of size $2$ of label $3$.
It is straightforward to check that all such single cluster of defects
will have leading weight $z^{-1}$.

The perturbation
expansion is well-defined in terms of number of
clusters~\cite{bellemans_nigam1nn,ramola}.
Thus, we write
\be\label{part2}
\frac{\mathcal{L}(z_o, z_e)}{\mathcal{L}(z_o,0)}=1+ W_1(z_o,z_e)+ 
W_2(z_o,z_e)+ \ldots,
\ee
where $W_n$ represents the contribution from $n$ clusters. 
The free energy  may then written as a series:
\be
f(z_o,z_e) = f_0 (z_o)+  f_1 (z_o,z_e)+f_2 (z_o,z_e) + \ldots,
\ee
where $f_i$ corresponds to the contribution from $i$ clusters. From
\eqref{free}, we immediately obtain
\begin{align}
f_0 (z_o) &= \frac{-1}{N} \ln \mathcal{L}(z_o,0), \label{f0defn}\\
f_1 (z_o,z_e) &= \frac{-1}{N} W_1(z_o,z_e), \label{f1defn}\\
f_2 (z_o,z_e) &= \frac{-1}{N} \left[W_2(z_o,z_e)-\frac{W_1(z_o,z_e)^2}{2} \right].\label{f2defn}
\end{align}
The free energies $f_0$, $f_1$ and $f_2$ are calculated in the following 
subsections.

\subsection{\label{sI}Calculation of $f_0$}

$f_0$ is the contribution to the free energy from configurations that
do not have defects. Then the heads of all the rectangles are in rows
with label $1$ and the configuration in a particular row is
independent of the configurations in other rows, Hence, we can write
\be
\label{omega1}
\mathcal{L}(z_o,0)=\left[\Omega_p(z_o,L)\right]^{L/m},
\ee
where $\Omega_p(z_o,L)$ is the grand canonical 
partition function of a system of hard
rods of length $d$ on a one dimensional lattice of $L$ sites  with periodic
boundary condition. The one dimensional partition function obeys
simple recursion relations.
$\Omega_p(z_o,L)$ is related to $\Omega_o(z_o,L)$, the corresponding
grand canonical partition function on a one dimensional  lattice of
$L$ sites with open boundary conditions, as
\be
\Omega_p(z,L)= d  z \Omega_o(z,L-d)+\Omega_o(z, L-1),~L \geq d.
\label{eq:recursive0}
\ee
$\Omega_o(z,\ell)$ 
obeys the following recursion relation:
\begin{align}
\Omega_o(z,\ell)&=z \Omega_o(z,\ell-d)+\Omega_o(z, \ell-1), ~\ell \geq
d, \label{eq:recursive1}\\
\Omega_o(z,\ell)&=1, ~ 0\leq \ell < d.
\label{eq:recursive2}
\end{align}

Equation~\eqref{eq:recursive1} is solved by the ansatz
$\Omega_o(z,\ell) = A \lambda^\ell$, where $A$ is a constant.
Substituting into~\eqref{eq:recursive1}, we obtain
\be
\label{poly}
\lambda^{d}-\lambda^{d-1}-z=0.
\ee
Let $\lambda_1$ denote the largest root of~\eqref{poly}.
For arbitrary $d$, $\lambda_1$ may be solved as a perturbation
series in inverse powers of $z^{-1}$. By examining a few terms in the
expansion, we find that the series solution of $\lambda_1$ has the
following form:
\be\label{l5}
\lambda_1(z)= 
z^{1/d}+\frac{1}{d}+ \sum_{n=1}^\infty \frac{z^{-n/d}}{d^{n+1} (n+1)!}
\prod_{j=1}^n (j d -n).
\ee
The free energy $f_0$ is related to the $\lambda_1$ as
\be
f_0 = -\frac{1}{m} \ln \lambda_1(z_0).
\ee
Thus,
\be
f_0=-\frac{\ln z_o}{d m}- \frac{z_o^{-1/d}}{d m} 
-\sum_{n=2}^\infty \frac{z_o^{-n/d}}{n! d^n  m }  \prod_{j=1}^{n-1} (j
d-n).
\label{eq:f0}
\ee
Note that $f_0$ is a series in integer powers of $z_o^{-1/d}$ rather
than the usual Mayer expansion that is in integer powers of $z_o^{-1}$.

It will turn out later that, to calculate the contribution
from configurations with two defect-clusters, we will need knowledge 
of the partition function
$\Omega_o(z,\ell)$ for all $\ell$ and not just for large $\ell$ when
only the largest root $\lambda_1$ of~\eqref{poly} contributes.
$\Omega_o(z,\ell)$ is a linear combination of the $d$ roots of
(\ref{poly}). Denoting the roots by $\lambda_n$, 
\be
\label{omegaoo}
\Omega_o(z,\ell)=\sum\limits_{n=1}^{d} c_n \lambda_n^{\ell},
\ee
where $c_n$ are constants to be determined from~\eqref{eq:recursive2}.
Substituting~\eqref{omegaoo} in~\eqref{eq:recursive2} we obtain 
\begin{equation}\label{vandermnd}
\begin{pmatrix}
1 & 1 & 1& \ldots&1\\
\lambda_1 & \lambda_2 & \lambda_3 &\ldots & \lambda_d\\
\lambda_1^2 & \lambda_2^2 & \lambda_3^2 &\ldots & \lambda_d^2\\
\ldots&\ldots&\ldots&\ldots& \ldots\\
\ldots&\ldots&\ldots&\ldots& \ldots\\
\lambda_1^{d-1} & \lambda_2^{d-1} & \lambda_3^{d-1} &\ldots &
\lambda_d^{d-1}
\end{pmatrix}
\begin{pmatrix} c_1\\c_2\\ \cdot \\ \cdot \\ \cdot \\c_{d} 
\end{pmatrix}=\begin{pmatrix} 1\\1\\ \cdot \\ \cdot \\ \cdot \\1 \end{pmatrix}.
\end{equation}
By inverting the Vandermonde 
matrix in~\eqref{vandermnd}, we obtain
\be\label{cn}
c_n=(-1)^{d-1}\frac{\prod \limits_{\substack {j=1\\j\neq n}}^{d} 
(\lambda_j -1)}{\prod \limits_{\substack {j=1\\j\neq n}}^{d} 
(\lambda_n-\lambda_j) },
\ee

To obtain $c_n$'s as a series in $z_o^{-1/d}$, we need to first obtain
all $\lambda_n$'s as a series in $z_o^{-1/d}$. From~\eqref{poly}, we
immediately notice that $|\lambda_n| = z^{1/d}$, $z \to \infty$. Let
\be
\label{l1}
\lambda_n=z^{1/d}e^{i \theta_n}.
\ee
Substituting into~\eqref{poly}, we obtain
\be
\label{cossin1}
e^{i \theta_n d} = 1+ z^{-1/d} e^{i \theta_n (d-1)}.
\ee
The perturbative solution to~\eqref{cossin1} is straightforward to
obtain:
\be
\label{l4}
\lambda_n=z^{1/d} e^{i \theta_n} + \frac{1}{d} + \frac{(d-1) 
e^{-i \theta_n}}{2 d^2} z^{-1/d}+O(z^{-2/d}),
\ee
where
\be\label{theta}
\theta_n=\frac{2(n-1) \pi}{d},\quad n=1,2,\ldots, d.
\ee
Using the expression for $c_n$'s in~\eqref{cn}, we obtain
\be\label{cn2}
c_n=\frac{1}{d}+\frac{(d-1) e^{-i 2 \pi (n-1)/d}}{d^2}z^{-1/d}+
\ldots,~n=1,\ldots,d.
\ee

\subsection{\label{sII}Calculation of $f_1$}

A single cluster of defects of size $n$ consists of $n$ rectangles placed one
directly above the other, keeping the long sides parallel, with the heads being
in rows with labels $2, \ldots, m$ (see text before~\eqref{part2} for a
more precise definition). 
Given a defect-cluster of size $n$, the occupation of exactly $n+1$ rows 
with label $1$ are affected.  The contribution to $W_1$ from a single defect
cluster of size $n$ is then
$\left[\frac{\Omega_o(z_o,L-d)}{\Omega_p(z_o,L)}\right]^{n+1}z_e^n$. 
The bottom left corner of the cluster has to be on a row with label $2$, 
and there are $N/m$ ways of choosing this lattice site. 
In addition, we
need to account for the number of ways $H(n)$ that a cluster of size $n$ 
may be split into sub-clusters with different labels. The number of ways of
distributing $n$ rectangles into $m-1$ sub-clusters, where the
sub-clusters are arranged in ascending order of labels,
is
\be
H(n) = {n+m-2 \choose m-2}.
\label{eq:combinatorial}
\ee
Thus, the contribution to $W_1$ from configurations with a 
single cluster of defects is
\begin{align}
\label{free4}
W_1(z_o,z_e) &= \frac{N}{m}\sum \limits_{n=1}^{\infty}
H(n) \left[ 
\frac{\Omega_o(z_o,L-d)}{\Omega_p(z_o,L)}\right]^{n+1} z_e^n,\\
&= \frac{N}{m}\frac{\Omega_o(z_o,L-d)}{\Omega_p(z_o,L)}
\left[ \frac{1}{\left[1-\frac{z_e\Omega_o(z_o,L-d)}
{\Omega_p(z_o,L)}\right]^{m-1}}-1\right].
\label{eq:W1}
\end{align}

$\Omega_p(z_o,L)$ may be expressed in terms of $\Omega_o(z_o,L)$ using
\eqref{eq:recursive0}:
\begin{align}
\label{approx}
\Omega_p(z_o,L)\:&=\: \Omega_o (z_o,L-d) \left[ d z_o +
\frac{\Omega_o(z_o, L-1)}
{ \Omega_o (z_o, L-d)} \right], \\
&\underset{L\to \infty}{=}\Omega_o (z_o,L-d) \left[ d z_o + \lambda_1^{d-1}\right] .
\label{apprper}\
\end{align}
where in the limit $L \to \infty$, we used $\Omega_o(z_o,\ell) \sim
\lambda_1^\ell$.
Substituting into~\eqref{eq:W1}, using~\eqref{f1defn}, 
setting $z_e=z_o=z$, and expanding
for large $z$, we obtain,

\begin{eqnarray}
f_1(z,z) &=&  \frac{-1}{m d z}
\left[\frac{d^{m-1}}{(d-1)^{m-1}}-1\right]
+ \frac{ 1}{ m d^2 z^{1+1/d}} 
\left[\frac{d^{m-1}(d+m-2)}{(d-1)^{m}}-1 \right] \nonumber \\
&&+ O\left(\frac{1}{z^{1+2/d}} \right).
\label{free6}
\end{eqnarray}
Thus single clusters of defects contribute at $O(z^{-1})$.


\subsection{\label{sIII} Calculation of $f_2$}

$f_2$ is the contribution to the free energy from two defect-clusters. 
There are four types of possible configurations:
\begin{enumerate}[(a)]
 \item Clusters that are separated by at least one row of label $1$
that may be occupied with rectangles independent of other rows.
 \item Clusters that intersect.
 \item Clusters that do not have any overlap in the $y$-direction, and
are not of type (a) or (b).
 \item Clusters that have some overlap in the 
$y$-direction, and are not of type (b).
\end{enumerate}
We calculate the contribution from each of these types in the
following subsections.

\subsubsection{\label{(a)} Type (a): Clusters that are separated from
each other}

The contribution of clusters of type (a) to the free energy 
is exactly 
cancelled by the contribution from the product of two single clusters
$W_1^2$ [which is $O(z^{-2})$].
Thus, there is no contribution to the free energy.

\subsubsection{\label{(b)} Type (b): Clusters that intersect}

When two clusters intersect, $W_2$ is exactly zero since such
configurations are forbidden. However, there is a contribution to
$f_2$ from such configurations through the terms $W_1^2$. Since
$W_1$ is $O(z^{-1})$, the contribution to $f_2$ is $O(z^{-2})$. 
Later, we will not be keeping terms of $O(z^{-2})$, and we therefore
neglect the contribution to $f_2$ from such configurations.

\subsubsection{\label{sIV} Type (c): Clusters that do not overlap in
the $y$-direction}

\begin{figure}
\centering
\includegraphics[width=\columnwidth]{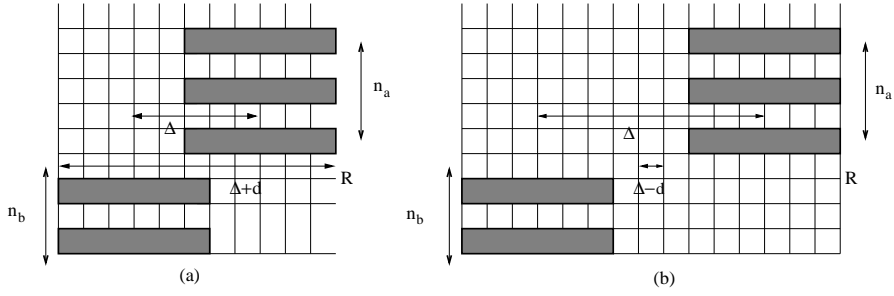}
\caption{Examples of configurations of two clusters with no overlap in
the $y$-direction. $\Delta$ is the distance between the centers of
the two clusters in the $x$-direction. Clusters could overlap in the
$x$-direction as in (a) $1 \leq \Delta <d$, or have no overlap in the
$x$-direction as in (b) $\Delta \geq d$. }
\label{fig:fig06}
\end{figure}

Let $W_2^x(z_o,z_e)$ denote the contribution to $W_2(z_o,z_e)$ from
configurations of clusters that have no overlap in the $y$-direction,
but may overlap in the $x$-direction. Examples of such configurations
are shown in Fig.~\ref{fig:fig06}. Let the number of defects in the
clusters be denoted by $n_a$ and $n_b$, and let 
$\Delta$ denote the distance in the $x$ direction between the centers
of the two clusters, where $\Delta \geq 1$. When $\Delta <d$, the two
clusters have some overlap in the $x$-direction [see
Fig.~\ref{fig:fig06}(a)], otherwise not [see 
Fig.~\ref{fig:fig06}(b)].

\begin{figure}
\centering
\includegraphics[width=\columnwidth]{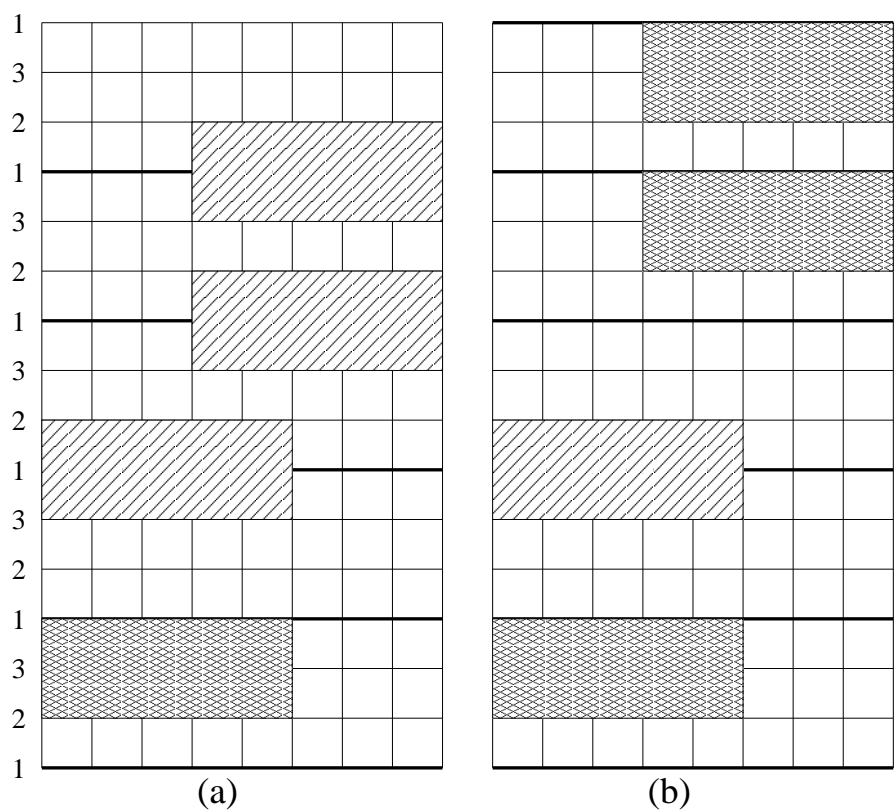}
\caption{Examples of configurations with two defect-clusters when $m=3$. 
(a) The lower cluster has two defects of labels $2$ and $3$ while
the upper cluster has two defects of label $3$. Such a configuration
affects the occupation of five rows of label $1$ (shown by bold lines). 
(b) The upper clusters has two defects of label $2$. Such a configuration
affects the occupation of $6$ rows of label $1$ (shown by bold lines).
 }
\label{fig:fig07}
\end{figure}

Let $t_1$ ($t_2$) denote the contribution to $W_2^x$ from pairs with
$\Delta<d$ ($\Delta \geq d$). We calculate $t_1$ and $t_2$ separately.
For $t_1$,  not all pairs of clusters with the same 
$\Delta$ contribute at the
same order in $z$. The lowest order contribution appears from pair of
clusters where the smallest label of the defects in the upper cluster
is larger than or equal to largest label of the defects in the lower
cluster. Other pairs of clusters contribute at $O(z^{-2})$. This may be
easily seen in the example shown in Fig~\ref{fig:fig07}, where two
cluster configurations are shown for the case $m=3$. In
Fig~\ref{fig:fig07}(a), the largest label of lower cluster is $3$ and the 
smallest label of the upper cluster is also $3$. Such a configuration of 
four defects affects five rows of label $1$ (denoted by bold lines)
and contributes at $O(z^{-1})$. In
Fig~\ref{fig:fig07}(b), the largest label of lower cluster is $3$ but the 
smallest label of the upper cluster is now $2$. Such a configuration of 
four defects affects six rows of label $1$ (denoted by bold lines), 
contributing at $O(z^{-2})$.

For the calculation of $t_1$, we will, therefore, restrict ourselves to cluster
configurations where the smallest label of the defects in the upper
cluster is larger than or equal to largest label of the defects in the lower 
cluster. In this case, the upper cluster may be slid to the left by
$\Delta$ till the two clusters merge to form a single cluster. Thus, the 
combinatorial factor
associated with dividing the two clusters into sub-clusters of
different labels is, as in the case of single clusters [see
Sec.~\ref{sII} and~\eqref{eq:combinatorial}], $H(n_a+n_b)$, 
where $n_a$ and $n_b$ are the
number of defects in the top and bottom clusters respectively. 
For $t_2$ ($\Delta \geq d$), the labels in each cluster are independent of
each other, and the combinatorial factor is therefore equal to $H(n_a)
H(n_b)$.
In addition, for both $t_1$ and $t_2$,
since we restrict $\Delta \geq 1$, a symmetry factor $2$ is associated
with each configuration. 

We now calculate the contribution to $t_1$ and $t_2$  from the occupation
of rows of label $1$ with rectangles.
The occupation of $n_a + n_b+1$ rows of
label 1 with rectangles are affected by the presence of the defects. The remaining
rows of label $1$ may be filled independently of each other. 
Among the $n_a + n_b+1$
rows, other than the row marked by $R$ in Fig.~\ref{fig:fig06}(a) and (b), $n_a+n_b$
rows may be thought of as open chains of length $L-d$. 
When $\Delta<d$ [see Fig.~\ref{fig:fig06}(a)], the row R is equivalent of an open
chain of length $L-d-\Delta$. When $\Delta \geq d$ [see Fig.~\ref{fig:fig06}(b)], the
row R is equivalent of two open chains of lengths $\Delta-d$ and $L-2d-(\Delta-d)$. 
Thus, we obtain
\be \label{t1}
t_1= \frac{2N}{m}\sum \limits_{n_a,n_b=1}^{\infty} 
\!\!H(n_a+n_b) 
\left[ \frac{\Omega_o(z_o,L-d) z_e}{\Omega_p(z_o,L)} \right]^{n_a+n_b}
\sum 
\limits_{\Delta = 1}^{d-1} 
\frac{\varOmega_o(z_o,L-d-\Delta)}{\varOmega_p(z_o, L)},
\ee
and
\begin{align}\label{t2}
t_2&= \frac{2N}{m}\sum \limits_{n_a,n_b=1}^{\infty} 
H(n_a)
H(n_b)
\left[ \frac{\Omega_o(z_o,L-d) z_e}{\Omega_p(z_o,L)} \right]^{n_a+n_b}
\nonumber \\
& \times \sum \limits_{\Delta = d}^{\infty} \frac{\Omega_o(z_o,L-d-\Delta)
\Omega_o(z_o,\Delta-d)}{\Omega_p(z_o,L)},
\end{align}
where the factor $N/m$ accounts for the number of ways of placing the
lower cluster on sublattice $2$.

In the limit $L\to \infty$, $\Omega_p(z_o,L)$ may be expressed in terms
of $\Omega_o(z_o,L)$ using~\eqref{apprper}. The limit of
large $L$ does not apply to the term 
$\Omega_o(z_o,\Delta-d)$, as $\Delta-d$ may be as small as zero. Hence, we
use~\eqref{omegaoo} for the one dimensional partition function for
any length. We, then, obtain
\be \label{q3a}
t_1= \frac{2N}{m}\sum \limits_{n_a,n_b=1}^{\infty} 
H(n_a+n_b) 
\left[\frac{z_e}{d z_o+ \lambda_1^{d-1}} \right]^{n_a+n_b}
 \sum \limits_{\Delta = 1}^{d-1} 
\frac{\lambda_1^{-\Delta}}{d z_o+ \lambda_1^{d-1}},
\ee
and 
\be \label{q3b}
t_2= \frac{2N}{m}\sum \limits_{n_a,n_b=1}^{\infty} 
H(n_a) H(n_b) 
\left[ \frac{z_e}{d z_o+ \lambda_1^{d-1}} \right]^{n_a+n_b}
\sum \limits_{\Delta = d}^{\infty} \sum \limits_{j = 1}^{d}
\frac{c_1 \beta_j \alpha_j^{\Delta-d} }
{\lambda_1^d (d z_o+ \lambda_1^{d-1})},
\ee
where  
\begin{align}
\label{betadef}
\beta_j &= \frac{c_j}{c_1},\\
\alpha_j&= \frac{\lambda_j}{\lambda_1},
\label{alphadef}
\end{align}
with $c_j$ and $\lambda_j$  as defined in~\eqref{omegaoo}.
Knowing the perturbation expansion for $c_j$ [see~\eqref{cn2}] and  
$\lambda_j$ [see~\eqref{l5} and
\eqref{l4}], the perturbation expansion for $\beta_j$ and $\alpha_j$
may be derived to be
\begin{align}
\beta_j&=1+\frac{d-1}{dz_o^{1/d}}\left[e^{-i 2\pi(j-1)/d}-1 \right] +
O(z_o^{-2/d}),~j=2,\ldots,d,
\label{beta}\\
\alpha_j&=e^{i\theta_j}+\frac{1}{dz_o^{1/d}}(1-e^{i\theta_j})+O(z_o^{-2/d}), ~j=2,\ldots,d,
\label{alpha}\\
\frac{1}{1-\alpha_j}&=\frac{1}{1-e^{i\theta_j}}
\left( 1+ \frac{1}{dz_o^{1/d}}\right) + O(z_o^{-2/d}),~j=2,\ldots,d,
\label{alpha2}
\end{align}
where $\theta_j$ is as defined in~\eqref{theta}. 

We now focus on $t_2$ [see~\eqref{q3b}]. 
Since $|\alpha_j|<1$ for $j>1$, the sum over
$\Delta$ reduces to a convergent geometric series for $j>1$.
Summing over $\Delta$, we obtain
\begin{align} \label{q4}
t_2&= \frac{2N}{m}\sum \limits_{n_a,n_b=1}^{\infty} 
H(n_a)
H(n_b)
\left( \frac{z_e}{d z_o+ \lambda_1^{d-1}} \right)^{n_a+n_b}
\nonumber\\
& \times 
\left[\sum \limits_{\Delta = d}^{\infty}  \frac{c_1  }
{\lambda_1^d (d z_o+ \lambda_1^{d-1}) }  +\sum \limits_{j = 2}^{d}
\frac{c_1 \beta_j  }
{\lambda_1^d (d z_o+ \lambda_1^{d-1}) (1-\alpha_j)} \right]
\end{align}
Though the first term in~\eqref{q4} is divergent, in the calculation for
$f_2$, it may be checked that this term is cancelled by the term $-W_1^2/2$. 
From~\eqref{cn2} $c_1 \sim d^{-1}$, and~\eqref{beta} $\beta_j \sim O(1)$. Similarly $|\alpha_j|
\sim 1$, though $\alpha _j \neq 1$ for $j>1$ [see~\eqref{alpha}]. Also $\lambda_1 \sim z^{1/d}$
[see~\eqref{l5}]. Therefore, we obtain that the second term in~\eqref{q4} is 
$O(z_o^{-2})$. Therefore, we conclude that  $t_2 \sim O(z^{-2})$. 

In the expression for $t_1$ [see~\eqref{q3a}], the leading order of the term
for a fixed $\Delta$ is $z^{1+\Delta}$. Keeping only the $\Delta=1$ term, 
we obtain 
\be\label{q5}
W_2^x=\frac{2N}{m}\sum \limits_{n_a,n_b=1}^{\infty} 
\frac{H(n_a+n_b)}{dz_o^{1+1/d}} 
\left(\frac{z_e}{dz_o}\right)^{n_a+n_b} +O(z^{-1-2/d}).
\ee
Doing the summation and setting $z_e=z_o=z$, we obtain
\be\label{q5a}
W_2^x=\frac{2 N}{m d z^{1+1/d}}
\left[1-\frac{d^{m-1}(d-m)}{(d-1)^m}\right] +O(z^{-1-2/d}).
\ee

\subsubsection{\label{sIIIa} Type (d): Clusters that overlap in
the $y$-direction}

\begin{figure}
\centering
\includegraphics[width=\columnwidth]{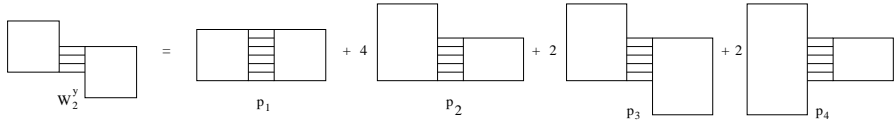}
\caption{Four possible cases of two clusters with overlap along $y$. In $p_1$,
both clusters do not extend beyond the common section. In $p_2$,
one of the clusters extends beyond the common section in one direction.
In $p_3$, both clusters extend beyond the common section. In $p_4$,
one of the clusters extends beyond the common section in both directions.
The numbers come from symmetry considerations and is straightforward
to obtain.}
\label{fig:fig08}
\end{figure}

Two non-intersecting rectangles of different labels are said to overlap in the
$y$-direction if the $y$-coordinate of the head of the rectangle with
larger label lies within the $y$-range 
of the rectangle with smaller label. Two rectangles of same label are
said to overlap in the $y$-direction if their heads are on the same row. Two
defect-clusters are said to overlap in the $y$-direction if they
contain at least one pair of overlapping rectangles.
Let $W_2^y$ denote the contribution from configurations with two
defect-clusters that have some overlap in the $y$-direction. 
We divide such configurations into four types, as shown in
Fig.~\ref{fig:fig08}, with their contributions to $W_2^y$ being
denoted by $p_1$, $p_2$, $p_3$ and $p_4$.
In $p_1$, neither of the clusters has any 
extension beyond the common section. In $p_2$, one of the clusters extends
beyond the common section in one direction. In $p_3$, both clusters extend beyond 
the common section in mutually opposite directions, while in $p_4$, one cluster extends beyond the 
common section in both directions. Each of the two clusters are single
clusters as defined in~\ref{sII} (also see Fig.~\ref{fig:fig04}). 
The configurations of type $p_2$, $p_3$ and $p_4$ may occur in four, 
two and two ways respectively, depending on which of the clusters is 
extending beyond the common section and in which direction. 

\begin{figure}
\centering
\includegraphics[width=0.9 \columnwidth]{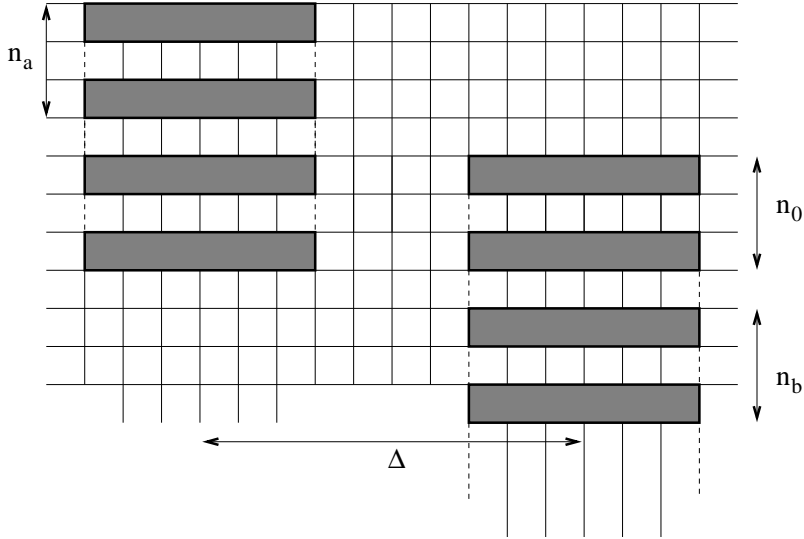}
\caption{An example ($m=2$, $d=7$) illustrating the definition of $n_0$, $n_a$,
$n_b$, and $\Delta$. $n_0$ is the number of rectangles in the common
section ($2$ in example),  
$n_a$ and $n_b$ are the extensions above  and below  ($2$ each) and
$\Delta$ is the horizontal distance between the centers of the two clusters
($10$ in example). }
\label{fig:fig09}
\end{figure}

For a pair of clusters, let $n_0$ denote the number of 
rectangles  that overlap in the $y$-direction.
Let the number of rectangles in the sections extending
above and below the common section be denoted by $n_a$ and $n_b$ 
respectively. As in Sec.~\ref{sIV}, let $\Delta$ be the horizontal distance 
between the 
centers of the two clusters. Clearly $\Delta\geq d$. These symbols
are illustrated in an example in Fig.~\ref{fig:fig09}. 

We now calculate $p_1$, $p_2$, $p_3$ and $p_4$. 
For $p_1$, the presence of the defect-clusters affects the 
occupation of exactly $(n_0+1)$ rows
of label $1$. The occupation of each of these rows with rectangles
is equivalent to
occupying two open chains of length
$\Delta-d$ and $L-d-\Delta$. Also, for each of the clusters, the
number of ways of breaking it up into sub-clusters  is
$H(n_0)$. Thus,
\be\label{p1}
p_1=\frac{N}{m} \sum \limits_{n_0=1}^{\infty} \sum 
\limits_{\Delta = d}^{\infty} H(n_0)^2
\left[\frac{\varOmega_o(z_o,L-d-\Delta)\varOmega_o(z_o,\Delta\!- \! d)}
{\varOmega_p(z_o, L)}\right]^{n_0+1} \! \! \! z_e^{2n_0}.
\ee
Expressing $\Omega_p$ in terms of $\Omega_o$ using~\eqref{apprper},
and $\Omega_o(z_o,\Delta- d)$ in terms of $\lambda_j$ using
\eqref{omegaoo}, we obtain
\be\label{p12}
p_1 =\frac{N}{m}
\sum \limits_{n_0=1}^{\infty}
\sum \limits_{\Delta = d}^{\infty} 
H(n_0)^2
\left[ \frac{c_1 \sum \limits_{\substack{j=1 }}^d \beta_j
\alpha_j^{\Delta-d}}{ \lambda_1^d (dz_o+\lambda_1^{d-1}) }
\right]^{n_0+1},
\ee
where $\beta_j$ and $\alpha_j$ are as defined in~\eqref{beta} and
\eqref{alpha}. Expanding the last term in~\eqref{p12} using
multinomial expansion, 
we obtain 
\begin{align}
p_1& =\frac{N}{m}\sum \limits_{n_0=1}^{\infty} H(n_0)^2 
z_e^{2n_0}\left[ \frac{c_1}{\lambda_1^d (dz_o+\lambda_1^{d-1})}\right]^{n_0+1}
\nonumber \\
& \times 
\sideset{}{'} \sum \limits_{\{n_l\}\geq 0 }
\frac{(n_0+1)!}{\prod \limits_{j=1}^d n_j!} \sum \limits_{\Delta= d}^{\infty}
\prod \limits_{p=2}^d (\beta_p \alpha_p^{\Delta-d})^{n_p},
\label{p13}
\end{align}
where the primed sum refers to the constraint
\be\label{constraint}
\sum \limits_{i=1}^{d}n_i=n_0+1.
\ee
Summing over $\Delta$, we obtain
\be\label{p14}
p_1 =\frac{N}{m}\sum \limits_{n_0=1}^{\infty}H(n_0)^2 
z_e^{2n_0}\left[ \frac{c_1 \lambda_1^{-d}}{ dz_o+\lambda_1^{d-1}}\right]^{n_0+1}
\sideset{}{'} \sum \limits_{\{n_l\}\geq 0 }
\frac{(n_0+1)! \prod \limits_{p=2}^d \beta_p^{n_p} }
{\prod \limits_{s=1}^d n_s! [1-\prod \limits_{j=2}^d \alpha_j^{n_j}]}.
\ee

We now estimate the order of the different terms in~\eqref{p14}.
From~\eqref{cn2} $c_1 \sim 1/d$ and from~\eqref{l4}, 
$\lambda_1^d\sim z_o$. Hence
\be\label{c1lambda}
\frac{c_1 \lambda_1^{-d}}{dz_o+\lambda_1^{d-1}}\sim \frac{1}{(dz_o)^2}.
\ee
To leading order, $\beta_i=1$, for all $i$ [see~\eqref{beta}]. Hence, unless
the term $1-\prod_{j=2}^d \alpha_j^{n_j}$ in \eqref{p14},
goes to zero as $z \to \infty$, the summand is
$O(z^{-2})$. Since we are not interested in terms of $O(z^{-2})$, we focus only
on those $\{n_l\}$ for which $\prod_{j=2}^d \alpha_j^{n_j} \approx 1$,
when $z \to \infty$. 
We obtain 
from~\eqref{alpha}
\be
\prod \limits_{j=2}^d \alpha_j^{n_j}= e^{\sum \limits_{ j=2}^d i \theta_j n_j}
\left[ 1-\sum \limits_{j=2}^d \frac{(1-e^{-i\theta_j})  n_j}
{dz_o^{1/d}} \right] + O(z_0^{-2/d}).
\ee
We are interested in those $\{n_l\}$ for which the leading term of the product
is $1$. This leads to the constraint
\be\label{2ndconstarint}
\sum \limits_{j=2}^d \theta_j n_j=2 l \pi, \quad l \in \mathbb{Z},
\ee
such that,
\be \label{prod1}
1-\prod \limits_{j=2}^d \alpha_j^{n_j}=\sum \limits_{j=2}^d 
\frac{(1-e^{-i \theta_j}) n_j}{d z_o^{1/d}}.
\ee

Let
\be
I(n_0) = \sideset{}{''} \sum \limits_{\{n_l\} \geq 0}
\frac{1}{\prod \limits_{j=1}^d n_j!\sum \limits_{j=2}^d(1-e^{-i\theta_j})n_j},
\label{sumj}
\ee
where the double prime refers to the constraints~\eqref{constraint} and
\eqref{2ndconstarint}. It is straightforward to see that $I(n_0)$ is real.
$\alpha_j$'s appear as complex conjugate pairs. For every $\{n_l\}$
satisfying the constraints, there is a $\{n'_l\}$ obtained by interchanging the
$n_l$'s of all complex conjugate pairs, the corresponding summand
being the complex conjugate. Thus $I(n_0)$ is real.
Equation~\eqref{p14} then reduces to 
\be\label{p15}
p_1=\frac{N}{m}\sum \limits_{n_0=1}^{\infty}H(n_0)^2 
\left( \frac{z_e}{ dz_o}\right)^{2n_0}\frac{(n_0+1)! I(n_0)}
{d z_o^{2-1/d}} +O(z^{-2}).
\ee
Thus $p_1$ is of order $z^{2-1/d}$ and for $d>2$ does not contribute to order
$z^{1+1/d}$.

For configurations of type $p_2$, one of the clusters extends beyond the other, 
the two clusters being of length $n_o$ and $n_0 + n_a$, where
the common section in the $y$-direction has $n_o$ rectangles. The
presence of these defect-clusters affects the occupation of
$(n_o+n_a+1)$ rows of label $1$. Out of these rows, $n_o+1$ of them
are equivalent to open chains of length $\Delta-d$ and $L-d-\Delta$,
and the remaining $n_a$ are equivalent to an open chain of length
$L-d$. The number of ways of dividing the clusters
into sub-clusters is
$H(n_0+n_a) H(n_0)$. In addition, there is a factor of $4 N/m$
from symmetry considerations. Thus,
\begin{align}\label{p21}
p_2&=\frac{4N}{m}
\sum \limits_{n_0,n_a=1}^{\infty} 
\sum \limits_{\Delta= d}^{\infty} 
H(n_0)
H(n_0+n_a)
z_e^{2n_0+n_a} \nonumber \\
&\times 
\left[\frac{\varOmega_o(z_o,L-d)}{\varOmega_p(z_o, L)}\right]^{n_a}
\left[\frac{\varOmega_o(z_o,L-d-\Delta)\varOmega_o(z_o,\Delta-d)}
{\varOmega_p(z_o, L)}\right]^{n_0+1}.
\end{align}

For configurations of type $p_3$, both clusters extend beyond the common 
section in the $y$-direction,
one cluster being of length $n_0+n_a$ and the other being of length $n_0+n_b$.
The number of ways of dividing the clusters
into sub-clusters is $H(n_0+n_a) H(n_0+n_b)$. For configurations of type $p_4$, one 
cluster extends beyond the common section in both directions, 
one cluster being of length $n_0+n_a+n_b$ and the other being of length $n_0$.
The number of ways of dividing the clusters
into sub-clusters is $H(n_0+n_a+n_b) H(n_0)$. For both $p_3$ and $p_4$
the presence of the defect-clusters affects the occupation of
$(n_o+n_a+n_b+1)$ rows of label $1$. Out of these rows, $n_o+1$ of them
are equivalent to open chains of length $\Delta-d$ and $L-d-\Delta$,
and the remaining $n_a+n_b$ are equivalent to an open chain of length
$L-d$.  Each of these two types also has an 
additional factor of $2 N/m$ due to symmetry considerations. Thus, we
obtain
\begin{align}
p_3&=\frac{2N}{m}
\sum \limits_{n_a=1}^{\infty}  
\sum \limits_{n_b=1}^{\infty}
\sum \limits_{\Delta = d}^{\infty} 
\sum \limits_{n_0=1}^{\infty}
H(n_0+n_a) H(n_0+n_b)
z_e^{2n_0+n_a+n_b}
\nonumber\\
&\times 
\left[\frac{\varOmega_o(z_o, L-d)}{\varOmega_p(z_o, L)}\right]^{n_a+n_b}
\left[\frac{\varOmega_o(z_o, L-d-\Delta)\varOmega_o(z_o, \Delta-d)}
{\varOmega_p(z_o, L)}\right]^{n_0+1}\label{p31},
\end{align}
and
\begin{align}
p_4&=\frac{2N}{m}
\sum \limits_{n_a=1}^{\infty}  
\sum \limits_{n_b=1}^{\infty}
\sum \limits_{\Delta = d}^{\infty} 
\sum \limits_{n_0=1}^{\infty}
H(n_0+n_a+n_b) H(n_0)
z_e^{2n_0+n_a+n_b}
\nonumber\\
&\times 
\left[\frac{\varOmega_o(z_o, L-d)}{\varOmega_p(z_o, L)}\right]^{n_a+n_b}
\left[\frac{\varOmega_o(z_o, L-d-\Delta) \varOmega_o(z_o, \Delta-d)}
{\varOmega_p(z_o, L)}\right]^{n_0+1}\label{p41}.
\end{align}

The expressions for $p_2$ [see~\eqref{p21}], 
$p_3$ [see~\eqref{p31}] and $p_4$ [see~\eqref{p41}] are similar to
that for $p_1$ [see~\eqref{p1}] except for the factor
$z_e^{n_a+n_b} [\varOmega_o(z_o, L-d)/\varOmega_p(z_o, L)]^{n_a+n_b}$.
However, this factor is $O(z^0)$. Hence, $p_2$, $p_3$, and $p_4$ are
also of $O(z^{-2+1/d})$. Combining the contributions from 
$p_1$, $p_2$, $p_3$  and $p_4$, we obtain
\begin{align}\label{p}
W_2^y&= \frac{N}{m}\left[ \sum  \limits_{n_0=0}^{\infty} H(n_0)^2 
+ 4 \sum  \limits_{n_0=0}^{\infty}\sum \limits_{n_a=1}^{\infty}
H(n_0) H(n_0+n_a) \left( \frac{z_e}{dz_o}\right)^{n_a} \right. \nonumber\\
&+ 2 \sum \limits_{n_0=0}^{\infty} \sum \limits_{n_a=1}^{\infty} 
\sum \limits_{n_b=1}^{\infty}
\left[H(n_0+n_a) H(n_0+n_b) + H(n_0+n_a+n_b) H(n_0)\right] 
\nonumber\\
&
\left. \left( \frac{z_e}{dz_o}\right)^{n_a+n_b} \right] 
\left( \frac{z_e}{dz_o}\right)^{2n_0}\frac{(n_0+1)! I(n_0) }{d z_o^{2-1/d}}
+O(z^{-2}).
\end{align}

\subsubsection{\label{s335} Expression for $f_2$}

The contribution to the free energy $f_2$ from clusters with two
configurations may now be computed. $W_2^x$ [see~\eqref{q5a}]
contributes at order $z^{-1-1/d}$. It is straightforward to argue that
configurations with three clusters will contribute at utmost order
$z^{-1-2/d}$, hence we truncate $f_2$ at order $z^{-1-1/d}$. 
The leading term of $W_2^y$ is $O(z^{-2+1/d})$, which for $d>2$ is
much smaller than $W_2^x$. Hence $W_2^y$ does not contribute to $f_2$
for $d>2$. When $m=2$ and $d=2$ we should 
include $O(z^{-2+1/d})$ term in $W_2^y$. This equals $3\ln (9/8)$, and
coincides with the high-activity expansion for $2\times 2$ hard
squares~\cite{ramola}. 
Substituting for $W_2^x$ from~\eqref{q5a}, we obtain
\be\label{q6}
f_2=
\frac{-2}{m dz^{1+1/d}} \left[1-\frac{d^{m-1}(d-m)}{(d-1)^m} + 6
\delta_{d,2} \ln \frac{9}{8} \right]
+O(z^{-1-2/d}).
\ee

\subsection{\label{s35} High activity expansion for the free energy
}

The free energy up to order $z^{-1-1/d}$ is obtained by summing $f_0$
[\eqref{eq:f0}], $f_1$ [\eqref{free6}], and $f_2$ [\eqref{q6}]:  

\begin{align}\label{free7}
-f(z,z)&=\frac{\ln z}{d m}+\frac{z^{-1/d}}{d m}
+\sum_{n=2}^{d+1} \frac{z^{-n/d}}{n! d^n  m }  \prod_{j=1}^{n-1} (j d-n) +
\frac{z^{-1}}{m d } \left(\kappa-1\right)\nonumber\\
&+ \frac{z^{-1-1/d}}{m d^2} \left[1+2d-\frac{\kappa\left(2 d^2 - 2 d
m+d+m-2 \right)}{d-1} +24 \delta_{d,2} \ln \frac{9}{8}\right] 
\nonumber \\
&+ O(z^{-1-2/d}).
\end{align}
where
\be
\kappa = \left(\frac{d}{d-1}\right)^{m-1}.
\label{kappa}
\ee

\section{\label{IV} Densities and transition points}

In this section we derive the high-activity expansion for the occupation 
densities of different rows. 
We truncate these expressions at $O(z^{-1-1/d})$ and then estimate the 
critical densities and activities for the nematic-columnar 
transition, and obtain their dependence on $m$ and $k$. 

Let $\rho_o$ ($\rho_e$) denote the number of lattice sites occupied by
rectangles whose heads are in rows with label $1$ (label different
from $1$). For rows with labels different from $1$, the occupation
densities will be equal. Hence,
\begin{align}
\rho_o=-m d z_o\frac{d}{ d z_o}f(z_o,z_e)\label{rho_o},\\
\rho_e=\frac{-m d z_e}{m-1}\frac{d}{ d z_e}f(z_o,z_e)\label{rho_e},
\end{align}
where the factor $m d$ accounts for the volume of a rectangle, and the
factor $m-1$ accounts for the $m-1$ labels that are different from $1$.
We, thus, obtain (after setting $z_o=z_e=z$)
\begin{align}
\rho_o&=1-\frac{z^{-1/d}}{d} -\sum \limits_{n=2}^{d+1} 
\frac{z^{-n/d} }{d^n (n-1)!}\prod_{j=1}^{n-1} (j d-n)+
\frac{1}{z}\left[ 1-\frac{\kappa (d+m-2)}{d-1}\right] \nonumber\\
&-\frac{(2d+1)(d+1)}{d^2 z^{1+1/d}} -\frac{\kappa}{(d-1)^2 z^{1+1/d}}
\Big[\frac{ m - 2}{d^2} 
- \frac{m^2 - m+1}{d} \nonumber\\
&   + 
2 m^2 - 5 m + 6 + (2 m - 1)d  - 2 d^2 +(17\ln\frac{9}{8}+4)\delta_{d,2} \Big] + O(z^{-1-2/d}), \label{rhoo}\\
\rho_e&=\kappa 
\left[\frac{1}{z (d-1)}+\frac{2md-2d-m+2+(16\ln\frac{9}{8}+8)\delta_{d,2} }
{d (d-1)^2 z^{1+1/d} }\right] + O(z^{-1-2/d}), \label{rhoe}
\end{align}
where $\kappa$ is as defined in \eqref{kappa}.

\begin{figure}
\centering
\includegraphics[width=0.9\columnwidth]{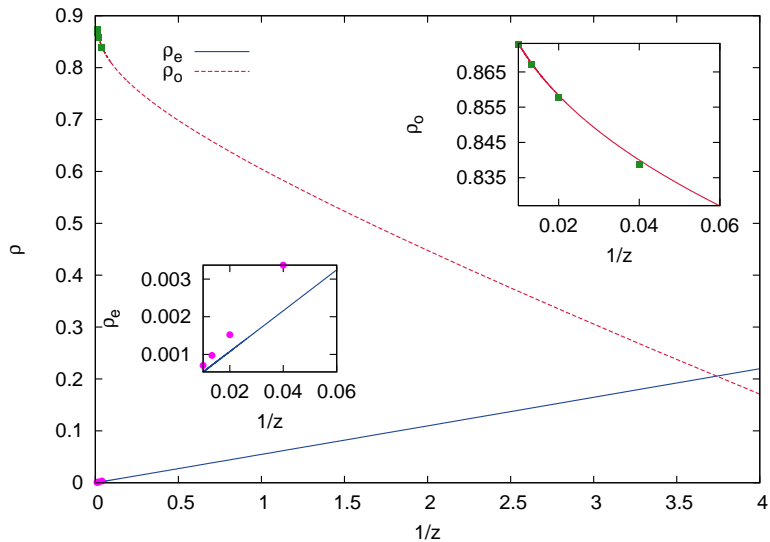}
\caption{The high activity expansions \eqref{rhoo} and \eqref{rhoe} for  $\rho_o$ 
and $\rho_e$ truncated at order $z^{-1-1/d}$ for 
$m=2$ and $d=22$. The data points are from Monte Carlo simulations. 
The two curves intersect at $z_c \approx 0.267$ and 
$\rho_c=\rho_o+\rho_e \approx 0.411$.
Right inset: Blow up of the large $z$ region for $\rho_o$.
Left inset: Blow up of the large $z$ region for $\rho_e$. 
}
\label{fig:fig10}
\end{figure}

In Fig.~\ref{fig:fig10} we plot \eqref{rhoo} and \eqref{rhoe},
truncated at $O(z^{-1-1/d})$ for $m=2$ and $d=22$. 
With increasing $z^{-1}$, $\rho_o$ decreases while $\rho_e$ increases. 
The intersection point of these two curves gives an estimate of the transition
point. In the example shown in Fig.~\ref{fig:fig10},
the estimates for the critical parameters are 
$z_c\approx 0.267$ and $\rho_c\approx 0.411$, where $z_c$ and $\rho_c$ are the
critical activity and critical density respectively.

For large $z$, the
expressions \eqref{rhoo} and \eqref{rhoe} are a good approximation to the 
actual densities and
reproduce the Monte Carlo results quite accurately.
The comparison with the Monte Carlo results is in shown in the insets of 
Fig.~\ref{fig:fig10} for both $\rho_o$ and $\rho_e$, and in 
Table~\ref{table:compare} for $\rho$. For $z\geq 25.0$, the 
expansion matches with the Monte Carlo results up to the third 
decimal place. This serves as an additional check for the correctness
of the calculations.
\begin{table}
\caption{\label{table:compare} Comparison of the results for density $\rho$
from the high-activity expansion $\rho^{exp}$ [see \eqref{rhoo} and \eqref{rhoe}]
with results from Monte Carlo
simulations $\rho^{sim}$ for four different values of the activity $z$.
The data are for a system of $2 \times 22$ rectangles.
}
\centering
\begin{tabularx}{0.32\textwidth}{ccc}
\hline\hline
z  & $\rho^{exp}$ & $\rho^{sim}$ \\ \hline\\
25.0 & 0.8422 & 0.8420 \\
50.0 & 0.8594 & 0.8593\\
75.0 & 0.8680 & 0.8680 \\
100.0 & 0.8736 & 0.8736 \\
\hline\hline
\end{tabularx}
\end{table}

By truncating the high-activity expansion, $\rho_o$ is overestimated
and $\rho_e$ is underestimated. In addition, for $m>3$, the nematic-columnar
transition is first order in nature ($\rho$ has a discontinuity), and $z_c$ will be larger than the value of $z$ for
which  $\rho_o=\rho_e$. Hence,  the estimate for $z_c$ that we obtain by setting 
$\rho_o=\rho_e$ in the truncated series is  a lower
bound to the actual $z_c$. For example, for $2 \times 2$ rectangles, $z_c$ increases
from $6.250$ when the series are truncated at $O(z^{-1})$ to
$14.859$ when the series are truncated  at $O(z^{-1-1/d})$ while the actual 
value is $97.5$~\cite{ramola}, while for $2 \times 22$ rectangles the 
corresponding values are $0.253$, $0.267$, and $2.11$ 
(see Fig.~\ref{fig:fig02}) respectively. We find that the increase 
in $z_c$, when terms of $O(z^{-1-1/d})$ 
are included, scales as $k^{-2}$, while the corresponding 
increase in $\rho_c$ scales as $k^{-1}$.

\begin{figure}
\centering
\includegraphics[width=0.9\columnwidth]{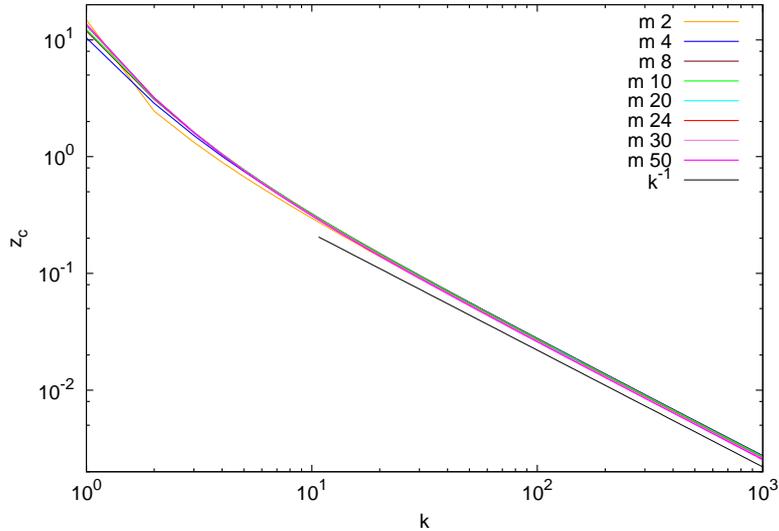}
\caption{Variation of $z_c$ with $k$ for different $m$. 
 }
\label{fig:fig11}
\end{figure}

We now study the dependence of the estimated $z_c$ and $\rho_c$ on $m$
and $d$. Figure~\ref{fig:fig11} shows the variation of $z_c$ with $k$ for
different values of $m$. For large $k$, $z_c$ decreases with $k$ as a power law
$c k^{-1}$, where $c\approx 2.78$ is independent of $m$. For small $k$, there is a 
crossover to a different behavior that depends weakly on $m$. We could 
not collapse the data for different $m$ onto a single curve by scaling $k$ with
a power of $m$. Hence, most likely the crossover scale increases logarithmically
with $m$. To summarize,
\be
z_c \approx \frac{2.78}{k}, \quad k \gg 1, \quad \forall m.
\ee
This behavior is in qualitative agreement with the result from Monte Carlo 
simulation where $z_c\approx35.5k^{-1}$ (see Fig.~\ref{fig:fig02}).

\begin{figure}
\centering
\includegraphics[width=0.9\columnwidth]{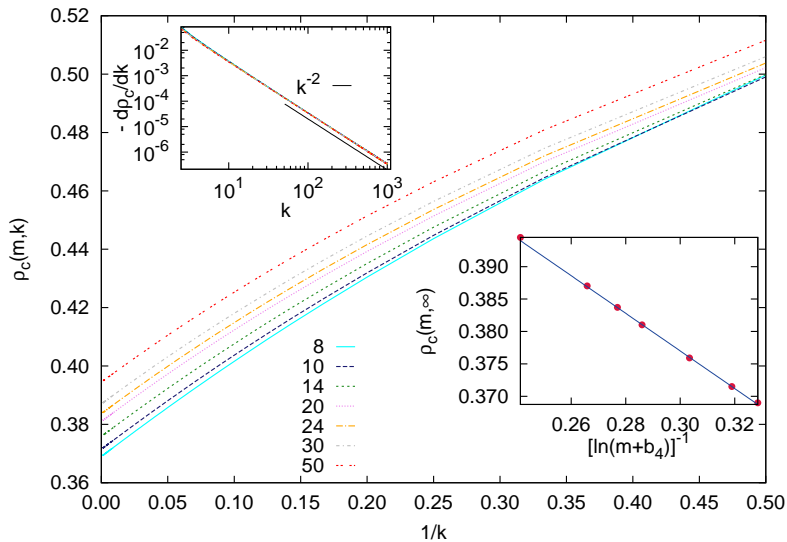}
\caption{Variation of $\rho_c(m,k)$ with $k$ for different $m$. 
Right inset: $\rho_c(m,\infty)$, the critical density for rectangles
with infinite aspect ratio, is well-described by
\eqref{rhocinfform} 
$b_2=0.464$, $b_3=0.29$, and $b_4=13.0$ (fit shown by solid line).
Left inset: The data for
$d\rho_c /dk$ for different $m$ collapse 
onto a single curve, independent of $m$.}
\label{fig:fig12}
\end{figure}

The variation of the corresponding critical density
$\rho_c(m,k)$ 
with $k$ and $m$ is shown in
Fig.~\ref{fig:fig12}. For large $k$, $\rho_c$ decreases linearly with $k^{-1}$ 
for all $m$. This dependence is brought out more clearly by 
$d\rho_c /dk= \rho_c(k+1)-
\rho_c(k)$ (see left inset of Fig.~\ref{fig:fig12}), which is
independent of $m$ and decreases as $k^{-2}$ for large $k$. We may
thus write
\be
\rho_c(m,k) = \rho_c(m,\infty) + \frac{b_1}{k}, \quad k \gg 1,
\label{rhocform}
\ee
where $\rho_c(m,\infty)$ is the $m$-dependent critical density for 
systems of rectangles of infinite aspect ratio, and $b_1\approx 0.367$
  is independent of $m$. For the corresponding data from Monte Carlo simulation $b_1\approx 0.226$ (see Fig.~\ref{fig:fig02}). We determine
$\rho_c(m,\infty)$
by fitting the data to \eqref{rhocform}, and its dependence on $m$ 
is shown in the
right inset of Fig.~\ref{fig:fig12}. We find that the data is
well-described by the form
\be\label{rhocinfform}
\rho_c(m,\infty) \approx b_2-\frac{b_3}{\ln (m+b_4)}, ~m \gg 1,
\ee
with $b_2 \approx 0.464$, $b_3 \approx 0.29$ and $b_4 \approx 13.0$.

Note that the $k$-dependence of $\rho_c(m,k)$ [see \eqref{rhocform}] is in 
qualitative agreement
with the results from Monte Carlo simulations for systems with $m=2$ (see
Fig.~\ref{fig:fig02} and \cite{kr14b})
and those from Bethe approximation for large $k$~\cite{kundu3}. Thus, though
the estimates that we obtain are lower bounds for the actual transition, we expect
that the truncated expansions give the correct qualitative trends for the
critical parameters.

\begin{figure}
\centering
\includegraphics[width=0.9\columnwidth]{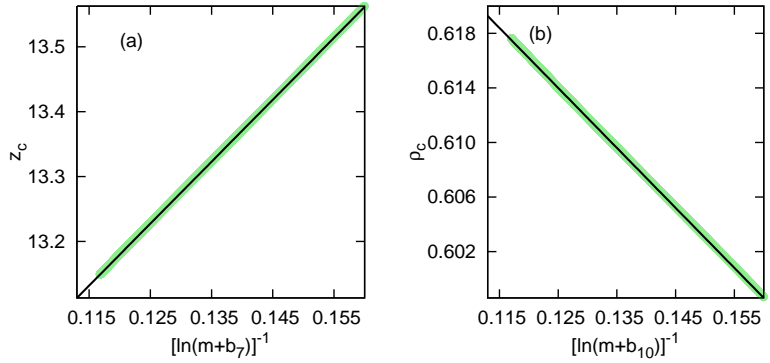}
\caption{Critical (a) activity $z_c$, and (b) density $\rho_c$, for 
a system of $m\times m$ hard squares. The straight lines correspond to (a)~\eqref{zcm1} with $b_5=12.085$, 
$b_6=9.54$ and $b_7=228.56$, and (b)~\eqref{rhocm1} with $b_8=0.669$, 
$b_9=0.44$ and $b_{10}=121.24$.
}
\label{fig:fig13}
\end{figure}

We now study the large $m$ behavior of the critical
density and critical activity of system of hard
squares of size $m\times m$, i.e., $k=1$. We find that $z_c$ increases
up to $m=97$ and then decreases to a constant for large $m$, while
$\rho_c(m,1)$ decreases up to $m=19$ and then increases to a constant
for large $m$. The asymptotic critical values are approached
logarithmically as the data are best described by (see
Fig.~\ref{fig:fig13})
\begin{align}
z_c(m,1) &=  b_5 + \frac{b_6}{\ln(m+b_7)},\label{zcm1}\\
 \rho_c(m,1) &=  b_8 - \frac{b_9}{\ln(m+b_{10})} ,\label{rhocm1}
\end{align}
with $b_5=12.085$, 
$b_6=9.54$, $b_7=228.56$, $b_8=0.669$, 
$b_9=0.44$, and $b_{10}=121.24$.

\section{\label{V}Conclusions and discussion}

In this paper, we derived the high-activity expansion for the free energy
and density of the columnar phase of $m\times d$ rectangles in a model where the
rectangles were restricted to be horizontal. The expansion is in inverse powers of
$z^{-1/d}$, where $z$ is the activity. We explicitly computed the first $d+2$ terms in this
expansion. As in the case for hard squares ($m=2$, $d=2$)~\cite{ramola}, the
expansion is not in terms of single defects, but in terms of clusters of defects. 

From the high-activity expansions for the densities of rectangles with heads
on rows of different labels, truncated at $O(z^{-1-1/d})$,
we estimate the transition points $z_c$ and $\rho_c$. We
show that $z_c \approx c k^{-1}$, $\rho_c(m,k) \approx \rho_c(m,\infty) + b_1k^{-1} $ for 
$k \gg 1$, where $c>0$, and $b_1>0$, are $m$-independent constants, 
and $\rho_c(m,\infty)$ 
increases logarithmically to a constant at large $m$.
For hard squares with $d=m$, or equivalently $k=1$, we obtain that the 
critical density increases logarithmically 
to a constant when $m \gg 1$. 

The high-density series being truncated at $O(z^{-1-1/d})$, the 
estimates for the critical parameters are lower bounds for
the actual values.
However, we note that the dependence of $\rho_c(m, k)$ and $z_c(m, k)$ on $k$ 
matches qualitatively  with the results obtained from Monte Carlo simulations 
for $m=2$ (see Fig.~\ref{fig:fig02} and \cite{kr14b}) and
Bethe approximation~\cite{kundu3} for large $k$. This leads us to conjecture 
that the trends for the critical parameters obtained from the truncated series
expansion are qualitatively correct. If that is true, the limit $m\to \infty$
keeping the aspect ratio $k$  fixed, corresponding to the limit 
of oriented rectangles in the continuum may be
studied. When $m \to \infty$, our results show that $z_c$ decreases to a
finite constant  for all $k$. If this feature carries over to the actual system,
then, it would imply that $\rho_c$ for large $m$ is less than one, implying that
the nematic-columnar transition should exist when $m \to \infty$. Likewise, there
should be a isotropic-columnar transition for hard squares ($k=1$) in the continuum at
a finite density. These conjectures should be verifiable using Monte Carlo for
systems in  the continuum.

The high density expansion derived in this paper was for a model where
all the rectangles were horizontal. This is a special case ($z_v=0$)
of the more general model where rectangles of horizontal and vertical
orientations occur with activity $z_h$ and $z_v$ respectively. We
argued, from Monte Carlo simulations, that setting $z_v=0$ does not
affect the nematic-columnar transition for large aspect ratio $k$.
However, the two models may differ for small $k$. For $k<7$, the
nematic phase does not exist when $z_v=z_h$~\cite{kundu3}. However,
when $z_v=0$, we have verified numerically that 
the nematic-columnar transition exists even for a system of
$2 \times 3$ rectangles. For $2 \times 3$ rectangles, when $z_v=z_h$, 
there are no density-driven phase transitions. Thus, in the two-dimensional
$z_v$--$z_h$ phase diagram, the phase boundary that originates at
$z_v=0$ must terminate on the line $z_h > z_v$, $z_v, z_h \to \infty$.
This leads to the interesting possibility that in the fully packed limit, 
as the ratio $z_h/z_v$ is increased, the system should undergo a
nematic-columnar transition. This conjecture should be verifiable for
systems like $2\times 3$ rectangles using Monte Carlo algorithms of
the kind recently implemented in Ref.~\cite{ramola3} where the fully
packed limit of mixtures of dimers and squares could be efficiently
simulated and shown to undergo a 
Berezinskii-Kosterlitz-Thouless
transition.

The high-activity expansion presented in the paper may be 
generalized to systems of polydispersed rods on lattices~\cite{i06,sr14} 
with same $m$ and different $d$. Suppose the rod-lengths are denoted
by $d_i$. The calculation of $f_0$ requires the knowledge of
only $\lambda$ [see \eqref{poly}]. The recursion relations
\eqref{eq:recursive0}, \eqref{eq:recursive1} and~\eqref{eq:recursive2}
obeyed by the one-dimensional partition functions will now be modified
to
\begin{align}
\Omega_p(\{z_i\},L) &= \sum_i d_i  z_i \Omega_o(\{z_i\},L-d_i)+
\Omega_o(\{z_i\}, L-1),~L \geq \text{max}[\{d_i \}],\\
\Omega_o(\{z_i\},\ell)&= \sum_i  z_i \Omega_o(\{z_i\},\ell-d_i)
\theta(\ell - d_i) +
\Omega_o(\{z_i\}, \ell-1),\nonumber \\
&~\ell \geq \text{min}[\{d_i \}], \\
\Omega_o(\{z_i\},\ell)&=1 ,~0\leq \ell < \text{min}[\{d_i \}],
\end{align}
where $z_i$ is the activity of a rod of length $d_i$, and $\theta(x)$
is the usual theta function. The
corresponding polynomial equation~\eqref{poly} is then modified to
\be\label{poly1}
\lambda^{d}-\lambda^{d-1}-\sum_i z_i=0,
\ee
making it possible to calculate the high-activity expansion for $f_0$.
The calculation of the 
contribution from configurations with a single cluster of defects is
also generalizable to the case of polydispersed rods. A defect cluster
now consists of rods of different lengths. The associated weight now
depends not just on the length of the cluster but also on the 
detailed structure of the cluster. Though more complicated, it is
still possible to write an exact expression for the contribution from
single defect-clusters.

Another possible extension of the derivation presented in this paper
is to find higher order correction terms. For example, suppose we
consider the $O(z^{-1-2/d})$ term. If $d \geq 4$, then there is no
contribution at this order from $W_2^y$ that accounts for defect
clusters with overlap along the $y$-direction. The contribution to
$O(z^{-1-2/d})$ term  from configurations with two or more
defect-clusters is only from $W^x$
that accounts for
defect-clusters with overlap in the $x$-direction.
This includes the $\Delta=2$ term in~\eqref{t1}, and contribution from
configurations with three defect-clusters, where each cluster is
misaligned from the bottom cluster by $\pm 1$. The corresponding term for 
three defect-clusters is
\be
\frac{4N}{m} \sum_{\{n_i\}} 
H\left[\sum_i n_i \right]
\left[ \frac{\Omega_o(z_o,L-d)z_e}{\Omega_p(z_o,L)} \right]^{\sum_i n_i -1}
z_e
\left[ \frac{\varOmega_o(z_o,L-d-1)}{\varOmega_p(z_o, L)} \right]^2,
\ee
where $n_i$ is the number of rectangles in defect-cluster $i$. It is
straightforward to check that the above expression contributes at
$O(z^{-1-2/d})$. Similarly, one can calculate the higher order 
terms up to $O(z^{2-2/d})$ by considering only $W^x$. 

Finally, it would be important to find an upper bound for the critical
parameters $z_c$ and $\rho_c$. This would amount to showing that that
the high-activity expansion derived in this paper has a finite radius
of convergence. However, this may not be easy as up till now, 
there exists no rigorous proof for the
existence of a columnar phase in any lattice model.

\section*{Acknowledgments}
The simulations were carried out on the supercomputing
machine Annapurna at The Institute of Mathematical Sciences.


\end{document}